# Slow Oscillations of the Transverse Magnetoresistance in HoTe$_3$

S. V. Zaitsev-Zotov[a,b,c,*], P. D. Grigoriev[d], D. Voropaev[a,c], A. A. Morocho[e],
I. A. Cohn[a,b], E. Pachoud[f], A. Hadj-Azzem[f], and P. Monceau[f]

[a] Institute of Radioengineering and Electronics, Russian Academy of Sciences, Moscow, 125009 Russia
[b] HSE University, Moscow, 101000 Russia
[c] Moscow Institute of Physics and Technology (National Research University), Dolgoprudnyi, Moscow region, 141700 Russia
[d] Landau Institute for Theoretical Physics, Russian Academy of Sciences, Chernogolovka, Moscow region, 142432 Russia
[e] National University of Science and Technology MISiS, Moscow, 119049 Russia
[f] Université Grenoble Alpes, CNRS, Grenoble INP, Institut NEEL, Grenoble, 38042 France
*e-mail: e-mail: serzz@cplire.ru

Slow oscillations of the magnetoresistance periodic in the inverse magnetic field with a frequency of 3.4 T have been identified in HoTe$_3$. The temperature dependence of the oscillation amplitude is close to exponential even at low temperatures. This may be attributed to the existence of soft modes in the system and allows the estimation of the electron scattering rate on these modes. In the region of magnetic fields exceeding 1 T, the oscillations can be described as interference oscillations associated with the splitting of the band structure due to the bilayer structure of HoTe$_3$. The obtained data have allowed us to calculate the ratio $2t_\perp/t_z = 15.6$ of the hopping integrals between layers ($t_\perp$) and between the bilayers ($t_z$) and to estimate these integrals as $t_\perp \sim 2$ meV and $t_z \sim 0.26$ meV.

## INTRODUCTION

The study of magnetic quantum oscillations is one of the most established and widely used methods for investigating the electronic structure of metals [1]. The frequency of these oscillations, when measured in the inverse magnetic field, is proportional to the cross-sectional area of the Fermi surface (FS). In ordinary metals, this frequency extends to thousands of tesla. As the range of studied substances was expanded, it turned out that some of them exhibited oscillations with extremely low frequencies of the order of a few tens and even units of tesla. Attempts to relate these oscillations to small FS pockets were unsuccessful due to the anomalously small size of the pockets and to a much weaker temperature dependence of slow oscillations compared to Shubnikov–de Haas oscillations.

There are several explanations for the origin of slow oscillations. All of these explanations are associated with the interference of two close frequencies of fast oscillations and are determined by the difference of their frequencies. The presence of slow oscillations has been observed in quasi-two-dimensional layered systems, and their occurrence is attributed to small interlayer electron hopping. This process leads to the formation of two extreme cross-sections of the cylindrical FS close in area due to its corrugation [2–4].

The occurrence of slow or difference oscillations is common in metals whose band structure has a splitting leading to the appearance of FS sites with close cross-sectional areas in the direction perpendicular to the direction of the magnetic field. Similar oscillations were previously studied in heterostructures and were called intersubband oscillations [5, 6]. Difference oscillations also occur in multiband conductors [7]. However, due to the variation in cyclotron frequencies across different bands, these oscillations frequently result in temperature decay [7], which aligns with the predictions of the Lifshitz–Kosevich formula [1], and that's what makes them different from slow oscillations in layered systems.

The splitting of the electronic spectrum is also observed in bilayer quasi-two-dimensional metals, which consist of two conducting atomic layers per unit cell. A prominent example of such material is high-temperature superconductors based on cuprates, where such slow oscillations have indeed been observed [8–14]. The origin of these oscillations remains a subject of debate, since their frequency is too low to arise from the FS calculated or observed in ARPES experiments. The origin of these low frequencies is ascribed either by FS rearrangement due to charge density waves (CDWs) [13, 14] or by the bilayer splitting of the electronic spectrum [15, 16]. The latter version is supported by the observed set of oscillation frequencies and angular dependence of these oscillation frequencies [12], and their weak dependence on the degree of doping [13].

In a single-particle approximation, when the thermodynamic potential is a linear functional of the electron density of states, slow oscillations of thermodynamic quantities such as the magnetization are strongly suppressed compared to slow oscillations of transport quantities such as the electronic conductivity [2, 3]. However, the electron–electron interaction gives rise to strong slow oscillations of thermodynamic quantities as well [17]. This probably explains the observation of slow quantum oscillations in cuprates not only in transport [8, 10–14] but also in thermodynamic quantities [9, 13, 14].

A situation similar to the cuprates has been observed in the rare-earth tritellurides ($RTe_3$), which also possess a bilayer crystal structure and CDWs. The occurrence of low-frequency magnetic quantum oscillations in these compounds has also been observed in [18], however, they also cannot be explained within the confines of the original FS. Below we will show that both mechanisms discussed for cuprate—the FS rearrangement due to CDW and the bilayer splitting of the electronic spectrum—occur in tritellurides. They give different frequencies of magnetic quantum oscillations, and the lowest frequency may be less than 10 T.

$HoTe_3$ is a representative of the tritellurides family of lanthanide group metals (R = La, Ce, Pr, Nd, Sm, Gd, Tb, Dy, Ho, Er, Tm) [19]. The materials of this family have an orthorhombic unit cell (space group *Cmcm*), comprising double Te planes separated by corrugated RTe planes. The $b$ axis is perpendicular to the Te planes, and the $HoTe_3$ unit cell consists of two Te–HoTe–HoTe–Te quadruple layers, which seems to form a next level-bilayer. These materials exhibit a complex phase diagram, characterized by the occurrence of two Peierls transitions as well as magnetic transitions at low temperatures [20].

A transition to a state with a CDW with a wave vector $\mathbf{Q}_{CDW1} = (0, 0, 0.296)$, which is incommensurate with the lattice is observed in $HoTe_3$ at a temperature $T_{P1} = 283$ K, while a second transition occurs at temperature $T_{P2} = 110$ K to form a CDW with a wave vector $\mathbf{Q}_{CDW2} = (0.32, 0, 0)$ perpendicular to $\mathbf{Q}_{CDW1}$. Nevertheless, at low temperatures, $HoTe_3$ retains the metallic conductivity, indicating the preservation of pockets on the FS after the two transitions with the formation of CDWs.

The magnetoresistance of $HoTe_3$ was investigated in a number of works [21, 22]. It has been revealed that a gradual transition from quadratic to linear magnetoresistance in the magnetic field occurs when the temperature is lowered from room temperature to 40 K, which is associated with the dependence of the electron scattering time $\tau_{hs}$ due to "hot spots" of FS on the magnetic field, $\tau_{hs} \propto 1/H$ [22]. Magnetoresistance oscillations in magnetic fields up to 65 T have also been studied [21]. It was revealed that in this compound, the magnetoresistance oscillation spectra depend on the range of investigation. Thus, in magnetic fields exceeding 40 T, magnetic breakdown occurs, accompanied by the emergence of high-frequency oscillation components with frequencies reaching up to 1700 T, while in magnetic fields up to 22 T, the frequencies do not exceed 100 T. Moreover, the spectra observed in magnetic fields of 3–16 and 6–22 T are significantly different. The characteristic oscillation frequencies observed in 3–22 T magnetic fields were attributed to the presence of small pockets on the FS in $HoTe_3$.

In this work, we study the lowest frequency component of the oscillations in $HoTe_3$. We reveal the evolution of low-frequency oscillations with a frequency of 3.4 T, starting from magnetic fields of 0.5 T in the temperature range from 3 to 60 K. The results demonstrate that the dependence of the oscillation amplitude on the inverse magnetic field is described by the predictions made in [16], where the influence of the splitting of the band structure on the quantum oscillations of the magnetoresistance of superconducting cuprates was considered.

## METHODS

A study was conducted using crystalline $HoTe_3$ samples grown by the gas transport method. The samples were thin rectangular plates with a thickness of approximately 30 μm and dimensions in the plate plane of 1–2 mm. High quality of the studied samples is confirmed by the substantial ratio $R(300 \text{ K})/R(4.2 \text{ K}) \approx 40$. The contacts to the samples were produced using a combined method with indium and silver paste, and were located at the corners of the samples. All measurements were conducted using an alternating current, with the orientation of the magnetic field perpendicular to the $a$–$c$ current flow plane.

## RESULTS

The temperature dependences of the resistance of $HoTe_3$ sample for two approximately orthogonal directions are shown in the inset of Fig. 1. The transition characterized by the formation of a high-temperature SDW at $T_{CDW1} = 285$ K is more prominent along one of the directions. This direction is henceforth referred to as $c$ direction, since it is in this direction that the transition to the conductivity tensor component exhibits its maximum effect. Accordingly, the other direction is regarded as the $a$ direction. The peculiarity of the temperature dependence of the conductivity, associated with the transition at $T_{\text{CDW2}} = 110$ K, is hardly manifested in conductivity measurements in the $ac$ plane [23].

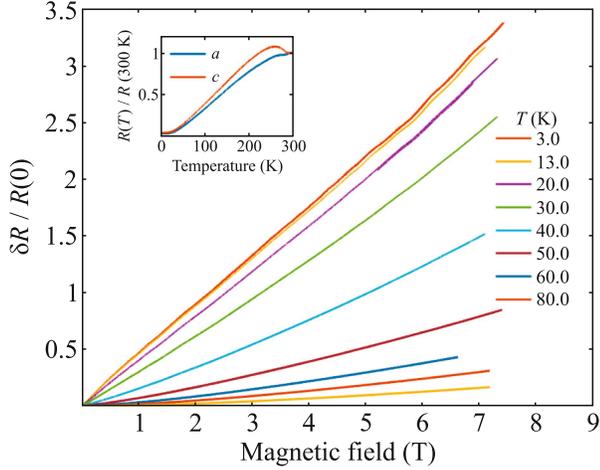

**Fig. 1.** (Color online) Temperature evolution of the magnetoresistance $\delta R/R(0)$ along the $c$ axis, where $\delta R = R(0) − R(B)$. The inset presents the temperature dependences of the sample resistance predominantly along the $a$ and $c$ directions.

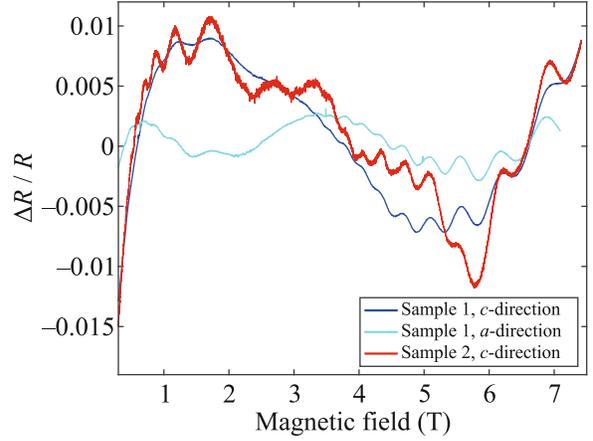

**Fig. 2.** (Color online) Magnetoresistance after the removal of the background approximated by a second-order polynomial. The measurement temperature is 3.0 K.

Figure 1 presents the dependences of the magnetoresistance along the $c$ axis on the magnetic field at temperatures of 3–80 K. All dependences are close to linear within the whole temperature range. The magnetoresistance along the $a$ axis as a function of the magnetic field exhibits a sublinear form that is only slightly pronounced (upward bending by approximately 10% in a field of 4 T). At the lowest temperatures, a weak oscillatory behavior manifests itself on the curves. Nevertheless, the magnetoresistance oscillations are small and are not perceptible on the original curves without further processing of the results.

The extraction of oscillations was conducted in two stages. At the first stage, a second-order polynomial was subtracted from the initial dependences $R(B)$, which made the oscillations more noticeable. Typical subtraction results for the two samples studied are shown in Fig. 2. It is evident that the oscillations can be categorized into at least two distinct groups: those observed in a low magnetic field below approximately 4 T and those observed in larger fields. The oscillations observed in low fields are the low-frequency oscillations of interest, while those observed in fields starting with approximately 4 T correspond to frequencies up to 60 T (see below), which are typical of HoTe$_3$ and have been studied previously [21]. It can also be seen that the low-frequency oscillations are noticeably larger in sample 2. The results presented below are obtained using this sample.

At the second stage, the data were replotted in the $\delta R/R(B)$ axis as a function of $1/B$, and the background curve obtained by applying sliding smoothing with a second-order polynomial in a window with a width of 0.6 1/T, was subtracted from the original data. This technique causes minor distortions in the amplitude of the oscillations near the ends of the investigated range, however, allows removing almost completely the slow background oscillations, clearly visible in Fig. 1, leaving just the oscillations studied in the present work. The result of this subtraction is shown in Fig. 3. The shape of the oscillations and their spectra measured in the $a$ and $c$ directions were almost identical.

The oscillation spectra measured at different temperatures are presented in Fig. 4. It is evident that oscillations with an frequency of 3.4 T predominate.

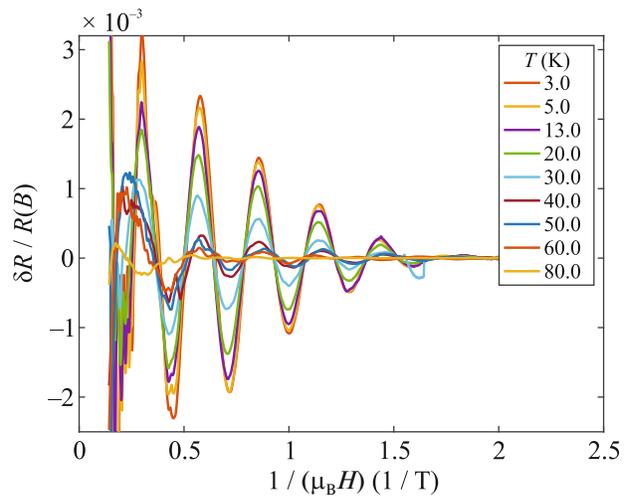

**Fig. 3.** (Color online) Oscillating part of the magnetoresistance.

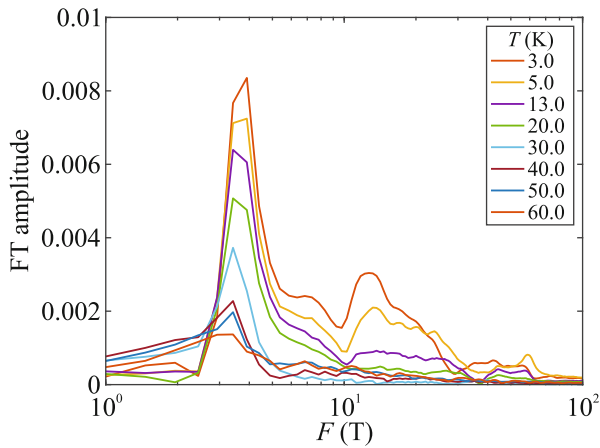

**Fig. 4.** (Color online) Oscillation spectra obtained from the data presented in Fig. 3 using the Fourier transform.

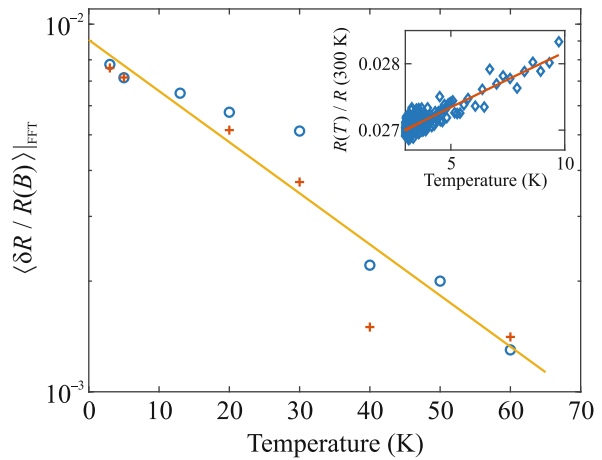

**Fig. 5.** (Color online) Temperature dependence of the amplitude of the low-frequency oscillation presented in Fig. 4. Different symbols show the amplitudes obtained from the data measured with the increase and decrease in the magnetic field. The inset presents the segment of the $R(T)$ dependence in the low temperature region. The line indicates data approximation by the second degree polynomial.

Figure 5 shows the temperature dependence of the oscillation amplitude obtained from the data presented in Fig. 4. Different symbols correspond to the results obtained with the increase and decrease in the magnetic field. Magnetic field hysteresis is observed, as evidenced by the data measured at one direction of the magnetic field sweep (circles), which are almost always higher than the data obtained at the opposite direction (crosses). Moreover, temperature hysteresis is observed as well. The data within the range of 3–30 K were taken at a consecutive temperature increases and lie mostly above the yellow line, while the data within the range of 40–70 K were taken at a decrease in temperature and predominantly fall below this line. The hysteresis loop is collapsed at the ends of the range. Within the limits of the error determined by the data spread and hysteresis, the dependence is close to exponential.

## DISCUSSION

As previously mentioned in the introduction, the observed linear magnetoresistance in materials exhibiting CDW, such as quasi-one-dimensional NbSe$_3$ [24], quasi-two-dimensional transition metal dichalcogenides, as well as TbTe$_3$ and HoTe$_3$ at elevated temperatures [22], was attributed to the dependence of the electron scattering time $\tau_{hs}$, resulting from the presence of FS hot spots on the applied magnetic field $\tau_{hs} \sim 1/H$ [22]. The findings of the present study suggest that the near-linear dependence observed in HoTe$_3$ persists also below 40 K. The detection of a 10% deviation from the linear dependence in a strong field for the $a$ direction (sublinear magnetoresistance) indicates the anisotropy of the FS in the CDW state and the competition of several electron scattering mechanisms, including hot spots, crystal defects, etc.

The detected oscillations of 3.4 T exhibit a frequency that is insufficient to be attributed to FS pockets, since the area of such a pocket would have to be $1.5 \times 10^{-4}$ of the Brillouin zone area (see supplementary materials).

The presence of bilayers in HoTe$_3$ inevitably leads to the splitting of the energy structure, as confirmed by the results of ab initio calculations [25, 26]. In our opinion, it is the presence of such splitting that leads to the appearance of low-frequency oscillations.

The electron dispersion relation along the interlayer direction for bilayer metals can be approximated by the formula [27]

$$\epsilon_{\pm}(k_z, \mathbf{k}_{\parallel}) = \epsilon_{\parallel}(\mathbf{k}_{\parallel}) \pm \sqrt{t_z^2 + t_{\perp}^2 + 2 t_z t_{\perp} \cos[k_z d]}, \quad (1)$$

where $t_{\perp}$ and $t_z$ are the integrals of electron hopping between layers within the bilayer and between the nearest bilayers, respectively, and $d$ is the lattice constant along the $z$ axis, equal to the distance between two identical layers of neighboring bilayers. Often, $t_z \ll t_{\perp}$ and the dispersion relation is simplified to the form

$$\epsilon_{\pm}(k_z, \mathbf{k}_{\parallel}) \approx \epsilon_{\parallel}(\mathbf{k}_{\parallel}) \pm t_{\perp}(\mathbf{k}_{\parallel}) \pm t_z(\mathbf{k}_{\parallel}) \cos[k_z d]. \quad (2)$$

The oscillations expected in the case of small splitting of the FS are described by the expression [16]

$$\sigma_2(\mu) \propto J_0^2\left(\frac{2\pi t_z}{\hbar \omega_c}\right) \cos\left(\frac{4\pi t_{\perp}}{\hbar \omega_c}\right) R_D^2, \quad (3)$$

where $J_0(x)$ is the Bessel function, $\omega_c = eB/m^*c\hbar$ is the cyclotron frequency, $m^*$ is the effective mass of carriers, and $e$ is the elementary charge.

According to the theory [2, 3, 16], Eq. (3) does not reflect the explicit temperature decay of slow oscillations. For ordinary quantum oscillations, this decay is expressed by the multiplier

$$R_T = R_T(T, B) = \frac{\lambda}{\sinh(\lambda)}, \quad \lambda \equiv \frac{2\pi k_B T}{\hbar \omega_c}. \quad (4)$$

Nevertheless, at sufficiently high temperatures the amplitude of slow oscillations decreases (see Fig. 5). We believe that this temperature decay arises due to the temperature dependence of the Dingle factor $R_D = \exp(-\pi/\omega_c \tau)$, which includes the scattering time $\tau$ of electrons on phonons, electrons, and other possible excitations. If these excitations are sufficiently soft, with energies below $k_B T$, their number increases linearly with the temperature, thereby yielding a linear temperature dependence of the scattering rate $1/\tau$. The existence of such excitations is substantiated by the observation of a nearly linear dependence $R(T)$ within the residual resistivity region (see inset in Fig. 5). This also leads to a near-exponential temperature dependence of the slow oscillation amplitude, shown by the orange line in Fig. 5, with $\tau \sim 10^{-11}$ s at $T = 3$ K. The hysteresis in the temperature dependence of the amplitude of magnetic quantum oscillations is related to the temperature hysteresis of the Hall effect known for HoTe$_3$ [28]. The hysteresis of this dependence on the magnetic field may be of the same nature.

It is known that at the lowest order of the electron–phonon interaction and for exponentially weak magnetic quantum oscillations, the Dingle factor $R_D$ and effective mass $m^*$ remain unchanged in the attenuation of magnetic quantum oscillation, given by equation (4) [29, 30]. This occurs due to a special reduction of two terms in the electron eigenenergy at $T \gg \hbar \omega_c$, which is included in both $R_D$ and $R_T$. This reduction was later confirmed for two-dimensional electron systems and for $e$–$e$ interactions [31–33], and was named the first-Matsubara-frequency rule [33]. The above mentioned reduction is derived for the temperature dependence of the amplitude of magnetic quantum oscillation [29–33], which contains the product of $R_T$ and $R_D$. The parameter $\tau$ itself or separately Dingle factor $R_D$ does not exhibit this reduction and depends on the temperature.

Soft modes in RTe$_3$, leading to temperature decay of slow oscillations even at $T < 30$ K, can arise due to CDW with nonideal nesting, when even at low temperature there are gapless electronic states at the Fermi level. Such fluctuations of the CDW lead to electron scattering in FS hot spots on the CDW wave vector, and have been discussed in connection with the linear dependence of the magnetoresistance in RTe$_3$ [22, 34]. The occurrence of soft modes in RTe$_3$ can be attributed to the competition between CDW and the crossing of bands at the Fermi level, resulting in hysteresis of the Hall coefficient [28]. The more pronounced decline in the amplitude of slow oscillations within the interval $T = 30$–$40$ K may be attributable to the incorporation of common phonon modes, which consequently results in a faster temperature dependence of the electron–phonon scattering rate $\tau^{-1} \sim T^3$ at $T$ significantly lower than the Debye temperature [1].

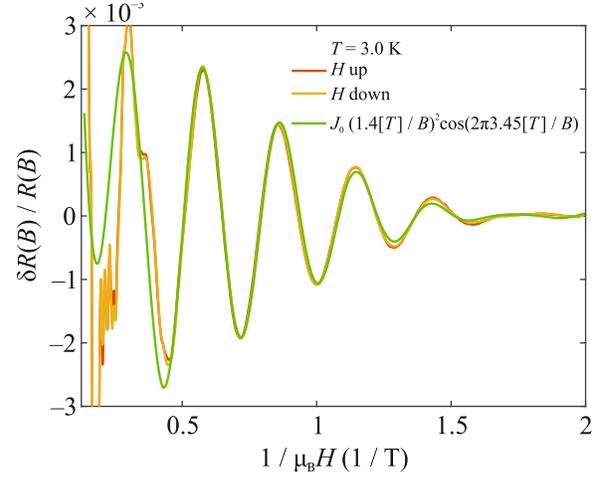

**Fig. 6.** (Color online) Experimental data at $T = 3.0$ K in comparison with the predictions made in [16].

Figure 6 illustrates the approximation of the measurement data by equation (3). The theoretical data underwent the same processing procedure as the experimental data, as described above. The resultant data demonstrate a high degree of congruence between the experimental results and the theoretical predictions.

Employing equation (3), from the obtained results we can calculate the ratio of hopping integrals $2t_\perp/t_z = 15.5$ and estimate their values from the relation $t_\perp = \frac{m_e}{m^*}\mu_B F$, where $F = 3.4$ T is the frequency of slow oscillations. To estimate the hopping integral values, it is necessary to know the effective mass of current carriers. For compounds belonging to the RTe$_3$ family that possess two CDWs, to which HoTe$_3$ belongs, the studies [21, 35] suggest the effective mass of carriers $m^*/m_e = 0.033$–$0.18$. Assuming $m^*/m_e = 0.1$ as a typical value, we obtain estimates of the hopping integrals $t_\perp \sim 2$ meV and $t_z \sim 0.26$ meV. The presence of such small hopping integrals is probably not indicative of the occurrence of HoTe–HoTe bilayers. Rather, it is more probable that these values are indicative of a much finer splitting of the band structure, arising due to the peculiarities of the crystal structure of HoTe$_3$, i.e., the existence of two Te–HoTe–HoTe–Te quadruple layers within the unit cell.

## CONCLUSIONS

To summarize, we have demonstrated the existence of slow oscillations of the conductivity of HoTe$_3$ in a magnetic field, obeying the predictions made [16]. The origin of these oscillations has been attributed to the bilayer structure of HoTe$_3$. These oscillations are almost entirely concentrated in the region of low magnetic fields (in the case of HoTe$_3$, $B \lesssim 4$ T), which is usually ignored when analyzing quantum oscillations of the conductivity. Thus, the study of oscillations in the region of low magnetic fields enables studying the fine splitting of energy bands.

## SUPPLEMENTARY INFORMATION

The online version contains supplementary material available at https://doi.org/10.1134/S0021364024605128.

## FUNDING

The work was supported jointly by the Russian Science Foundation (project no. 22-42-09018) and Agence nationale de la recherche (grant no. 21-CE30-0055).

## CONFLICT OF INTEREST

The authors of this work declare that they have no conflict of interest.

# Supplementary Material to the article
# "Slow oscillations of transverse magnetoresistance in HoTe$_3$"

## Supplementary Note 1

### Tight Binding model of RTe$_3$

The electronic band structure in the quasi-two-dimensional RTe$_3$ can be well approximated by the elementary tight binding (TB) model of the in-plane Te 5p$_{x/y}$ orbitals, which yields the following dispersions:

$$\epsilon_{px}(k_x, k_y) = -2t_{para}\cos[k_x a_0] - 2t_{perp}\cos[k_y a_0] - E_F$$

$$\epsilon_{py}(k_x, k_y) = -2t_{para}\cos[k_y a_0] - 2t_{perp}\cos[k_x a_0] - E_F \quad (1)$$

where $a_0 \approx c_0$ is the 2D in-plane lattice constant with magnitude √2 smaller than that of the 3D unit cell. The Fermi energy $E_F$ is determined from the electron density, namely from the condition of 1.25 electrons for each $p_x$ and $p_y$ orbitals [1]

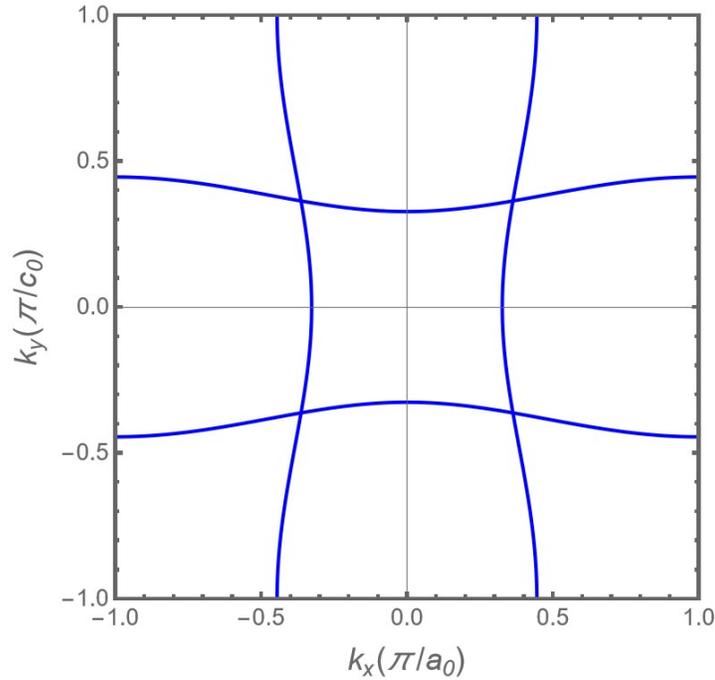

Fig. S1. 2D Fermi Surface (FS) of HoTe$_3$ calculated by TB model, using $t_{para}$ = −1.96 eV and $t_{perp}$ = 0.34 eV and $E_F$ = −1.35 eV [1].

The main (bare) bands of the non-interacting TB model, shown in Fig. S1, gain a small curvature proportional to the ratio $t_{\text{perp}}/t_{\text{para}}$ leading to the diamond-shaped Fermi surface. The CDW transition takes place when there is a permanent lattice distortion at some wave vector $\mathbf{Q}_{\text{CDW}}$. Taking the mean-field approach, the lattice distortion leads to a coupling between the states $|k\rangle$ and $|k \pm Q_{\text{CDW}}\rangle$ with an interaction strength of $\Delta$. Hence, the new electronic eigenstates $|\psi_k\rangle$ can be written as a superposition of the original states

$$|\psi_k\rangle = u_{k-Q_{\text{CDW}}} ¿$$

and are the eigenstates of the following matrix [2]:

$$\begin{bmatrix} \epsilon_{k-Q_{\text{CDW}}} & \Delta & 0 \\ \Delta & \epsilon_k & \Delta \\ 0 & \Delta & \epsilon_{k+Q_{\text{CDW}}} \end{bmatrix} \quad (3)$$

The eigenvalues of this matrix yield a new electronic dispersion, whose spectral weight $|u_k|^2$ is translated by $\pm \mathbf{Q}_{\text{CDW}}$ (shadow bands). This leads to the opening of a gap with amplitude of $2\Delta$ around the crossing of main $|k\rangle$ and shadow bands $|k \pm Q_{\text{CDW}}\rangle$.

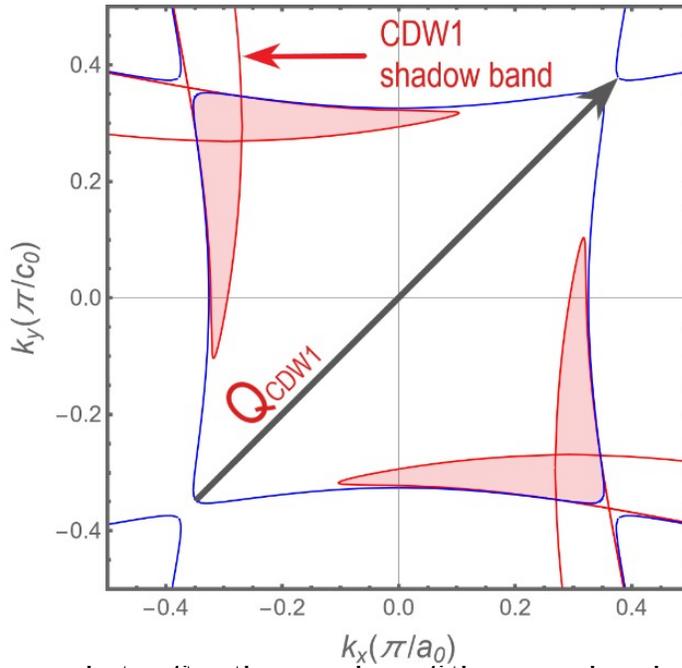

Fig. S2. Electron pockets after the overlap of the main bands $|k\rangle$ and shadow bands $|k \pm Q_{\text{CDW}_1}\rangle$. Due to imperfect nesting by the CDW$_1$ vector, the band overlap between

the states $|k\rangle$ and $|k\pm Q_{CDW1}\rangle$ and hence the position of the gap relative to $E_F$ shifts with momentum, which results in residual metallic pockets on the FS, marked with red color. Here, we emphasize the residual pockets inside the square part (blue square) of FS resulting from the interaction between $p_x$ and $p_y$. The coupling parameter $(\Delta_1=0.275/2\,eV)$ and $CDW_1$ vector of $HoTe_3$ can be extracted from [3].

Experimentally, the $CDW_1$ gap is well seen in [4], where the part near the Fermi level is emphasized. As we can see in [4, 5], with decreasing $k_y$ the nesting gradually weakens and residual metallic pockets appear on the FS, especially near their crossings, and therefore the system remains conducting in the CDW phase. Theoretically, the residual electron pockets (marked with red color) are shown in Fig. S2.

The first transition temperature $T_{CDW1}$ ranges from a low temperature of 244K for $TmTe_3$, the compound with the smallest lattice parameter in the $RTe_3$ series, and increases monotonically with increasing lattice parameter. In contrast, measurements for the smaller lattice parameter compounds R = Dy - Tm reveal a second $CDW_2$ feature at a lower temperature $T_{CDW2}$ ($T_{CDW2}$ is largest for the heaviest member of the series) [6]. Preliminary ARPES results for $ErTe_3$ confirm this picture, revealing additional gaps forming on sections of the FS close to the tips of the diamond sections of the FS pointing in $a^*$ direction [7]. The corresponding jump in the resistivity at $T_{CDW2}$, related to the amount of FS gapped at the transition, is largest for the compound with the largest value of $T_{CDW2}$ (smallest value of $T_{CDW1}$) [6] and smallest area of initial FS gapped at $T_{CDW1}$.

The effect of the perpendicular $CDW_2$ is approximated using the tight biding model and illustrated in Fig. S3. The crossing regions between the coupled bands $CDW_1$ and $CDW_2$ are marked with green (blue) color. The gaps due to both CDWs are denoted with dark rectangles. As it can be noticed, the spectral weight within the nested FS regions connected by the CDW vector $\mathbf{Q}_{CDW2}$ vanishes. This results in an amount of FS gapped at the transition $T_{CDW2}$ and hence in a reduction of the electron pockets as is shown in Fig. S3.

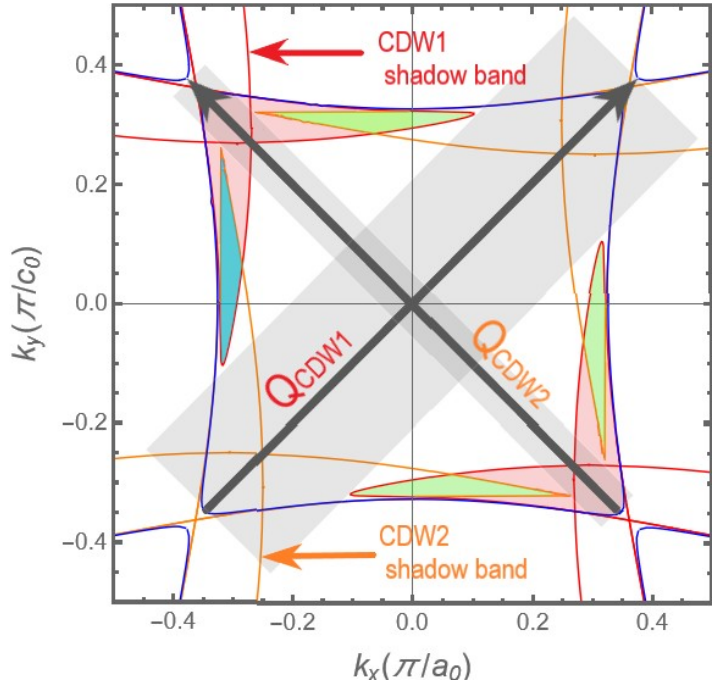

Fig. S3. Same as Fig. S2 but including the CDW$_2$ bands. The coupling parameter ($\Delta_2 = 0.142/2\,eV$) and CDW$_2$ vector of HoTe$_3$ can be extracted from [3]. The crossing regions marked with green (blue) color denote the electron pockets after the overlap of the CDW bands. Monte Carlo Method will be used to estimate the area of the blue colored region.

ARPES measurements for ErTe$_3$ at $T < T_{CDW2}$ show that the spectral weight is more intense at the crossing regions between the coupled bands CDW$_1$ and CDW$_2$ [7], which will be better seen in Supplementary Note 2.

Experimentally, the electron pockets resulting from the overlap of the main bands $|k\rangle$ and shadow bands $|k \pm Q_{CDW2}\rangle$ are not visible due to the large CDW$_1$ gap. This is well seen in Fig. S3.

## Supplementary Note 2

### Electron pocket area estimation by Monte Carlo Method

Since the positions of the pockets were determined by the spectral weight in ARPES [7], we select the crossing region marked with blue color in Fig. S3 and determine its area by Monte Carlo Method (MC), as is shown in Fig S4. The calculated Area is $A_{MC} = 0.67\%$ of the Brillouin zone (BZ).

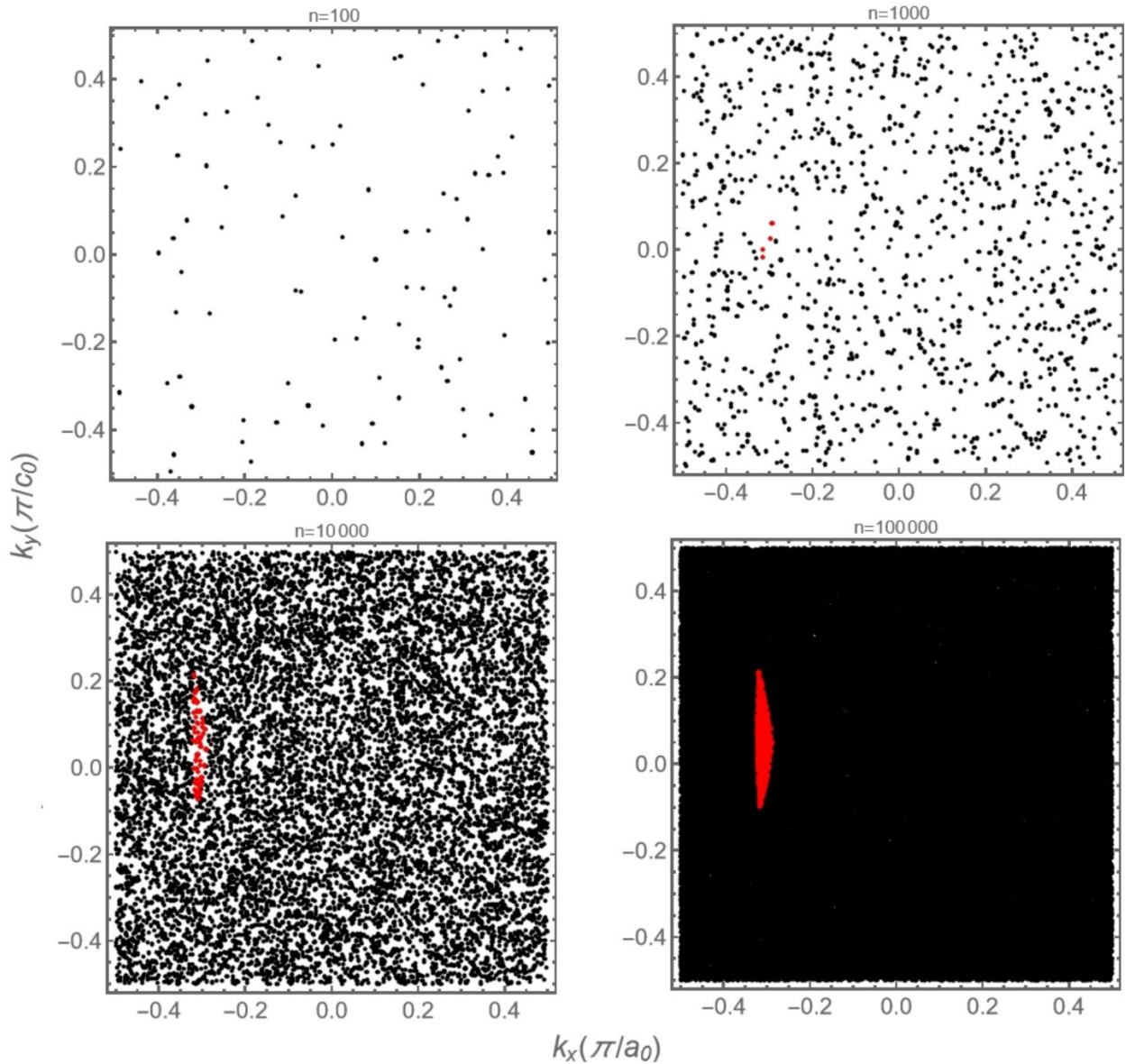

Fig. S4. MC area calculation of the region marked with blue color in Fig. S4. *n* is the total number of points inside the BZ.

## Electron pocket area estimation from ARPES

The area estimated by MC in Fig. S4 helps us to estimate the electron pocket areas previously observed in ARPES [7] (see Fig. S5). In this case, the position of the observed pockets is fixed by mapping their intensity in the spectral weight, and then we proceed to determine the number of covered pixels as well as the coverage of the colors. This method can give a good approximation of the area of the pockets since the area of the BZ is already known and covered with $n = 100000$ black points (see Fig. S4). This gives the following areas for the mapped regions **1** and **2**: $A_{AR1} = 0.64\%$ and $A_{AR2} = 0.55\%$, respectively.

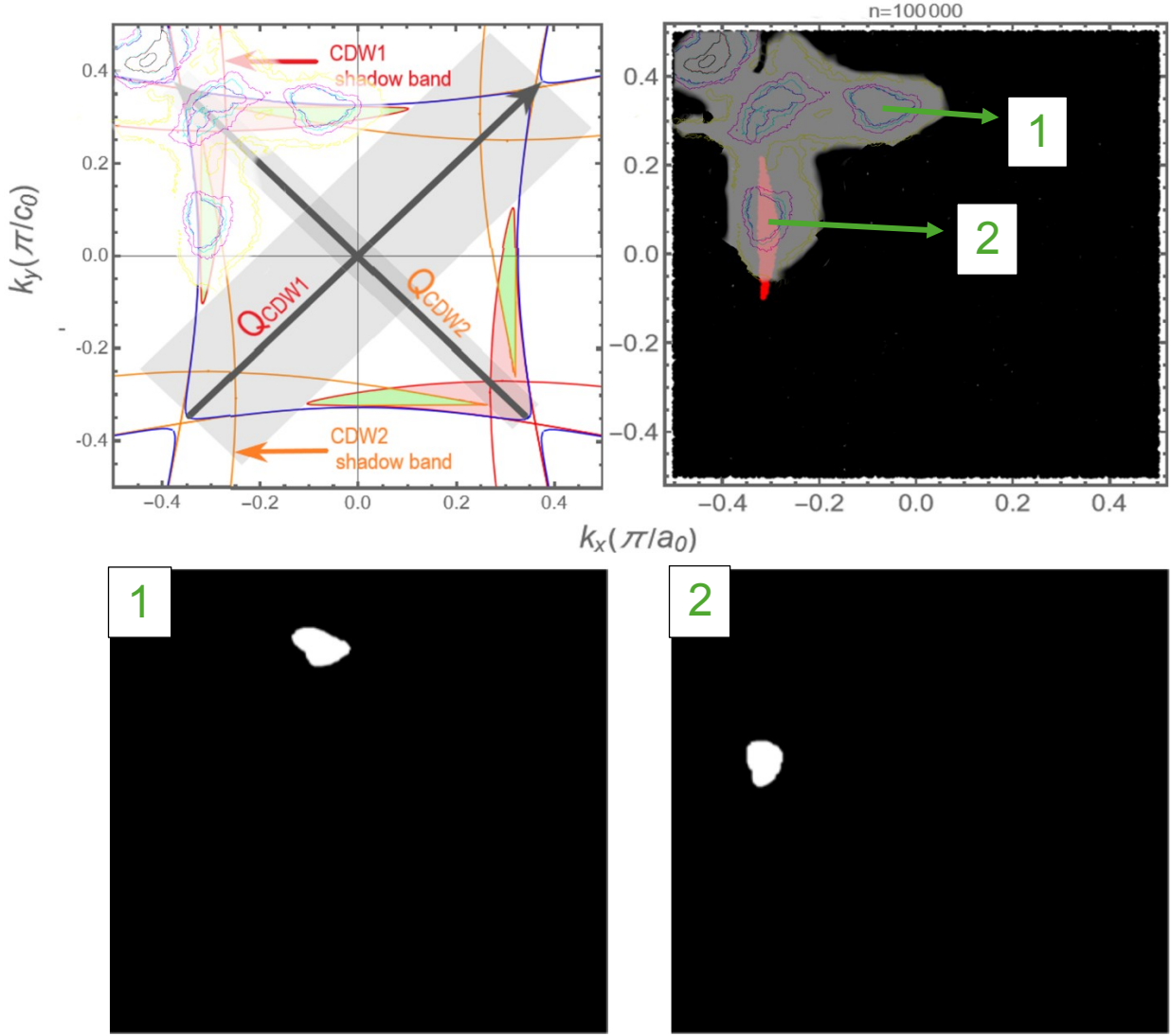

Fig. S5. Area determination of electron pockets observed in ARPES [7]. The spectral weight in ErTe$_3$ at $T < T_{CDW2}$ is fixed in the calculated bands and the location of the high-intensity regions is mapped for convenience.

As we can see from ARPES [7], the pockets shapes are not symmetrical, which is better observed in the mapped spectral weight from Fig. S5. This leads to different pocket areas in regions **1** and **2**. Otherwise, according to TBM, the pockets should have the same area after coupling. Hence, in addition to the inaccuracy of ARPES measurements, asymmetrical shapes of the pockets make it difficult to determine the areas.

# Supplementary Note 3

## Frequency determination

Quantum oscillations is a direct measure of the FS area via the Onsager relation:

$$F = \left(\frac{\phi_0}{2\pi^2}\right) A_H \quad (4)$$

where $\phi_0 = hc/2e = 2.07 \times 10^{-15}$ Tm$^2$ is the flux quantum, and $A_H$ is the cross-sectional area of the FS normal to the applied field [8]. This results in the following frequencies:

| Area | Frequency |
|---|---|
| $A_{MC} = 0.67\%$ | 151 T |
| $A_{AR1} = 0.64\%$ | 144 T |
| $A_{AR2} = 0.55\%$ | 124 T |

A frequency of 124 T implies a Fermi surface pocket that encloses a k-space area of $A_H = 0.55\%$ of the Brillouin zone.

The approximation can be improved by considering the region of intersection between $\mathbf{A_{MC}}$ and $\mathbf{A_{AR2}}$ ($A_{MC \cap AR2} = 0.3\%$) which gives a frequency of F = 67 T. Although the Onsager relation predicts lower areas for peaks observed in the quantum oscillation frequency spectrum in Fig. 4 of the main manuscript, the approximation given by us is very good despite the inaccuracy of the ARPES measurements.

It should be mentioned that the CDW gaps ($2\Delta_1$=175 meV and $2\Delta_2$=50 meV [7]) for ErTe$_3$ are smaller than the gaps for HoTe$_3$, and that the spectral weight distributed on the translated parts of the band structure strongly depends on the CDW vectors. This results in a larger margin of error in our calculation after the reconstruction of the FS in the TB model.